# Modelling ultrafast out-of-equilibrium carrier dynamics and relaxation processes upon irradiation of hexagonal Silicon-Carbide with femtosecond laser pulses

G. D. Tsibidis [1*], L. Mouchliadis [1], M.Pedio [2] and E. Stratakis [1,3]

[1] *Institute of Electronic Structure and Laser (IESL), Foundation for Research and Technology (FORTH), N. Plastira 100, Vassilika Vouton, 70013, Heraklion, Crete, Greece*

[2] *Istituto Officina dei Materiali, Consiglio Nazionale delle Ricerche (CNR-IOM), Trieste, Italy*

[3] *Department of Physics, University of Crete, 71003 Heraklion, Greece*

We present a theoretical investigation of the yet unexplored dynamics of the produced excited carriers upon irradiation of hexagonal Silicon Carbide (6H-SiC) with femtosecond laser pulses. To describe the ultrafast behaviour of laser induced out-of-equilibrium carriers, a real time simulation based on Density Functional Theory (DFT) methodology is used to compute both the hot carrier dynamics and transient change of the optical properties. A Two-Temperature model (TTM) is also employed to derive the relaxation processes (i.e. thermal equilibration between carrier and lattice through carrier-phonon coupling) for laser pulses of wavelength 401 nm, duration 50 fs at normal incidence irradiation which indicate that surface damage on the material occurs for fluence ~1.88 $Jcm^{-2}$. This approach of linking, for the first time, real time calculations, transient optical properties and TTM modelling, has strong implications for understanding both the ultrafast dynamics and processes of energy relaxation between carrier and phonon subsystems and providing a precise investigation of the impact of hot carrier population in surface damage mechanisms in solids.

## I. INTRODUCTION

Over the past decades, the advances of ultra-short pulsed laser technology have emerged as a powerful tool for many technological applications, in particular in industry and medicine [1-14]. To this end, understanding of laser driven physical phenomena such as electron excitation, scattering processes, relaxation mechanisms, phase transitions, and ablation are important to elucidate many fundamental properties of solids that can lead to enhanced control of the laser energy for numerous potential applications.

One of the most challenging issues that influences laser driven phenomena is the response of excited carriers scattering processes in the femtosecond time window. A better description of those mechanisms is crucial for a detailed knowledge of laser induced ultrafast processes. On the other hand, the investigation of the ultrafast electron dynamics within the electron gas in a laser-heated material is a real challenge. In principle, the extremely small electron-electron collision time (~10 fs), associated with the generation of highly hot and nonthermalised (i.e. out-of-equilibrium) electron distribution during excitation, complicate direct observation [15]. Nevertheless, advances in laser technology have allowed generation of out-of-equilibrium electron distributions while they have enabled observation of their relaxation in real time, through predominantly the response of the material optical parameters [16-18].

To model laser-matter interaction and describe material's response, a common approach that has been widely used is the traditional two temperature model (TTM) which, however, ignores the formation of *nonthermal electron* populations [19]. One major problem of the classical TTM is that it considers that these extremely hot excited carriers thermalize instantaneously which is not valid [20]. While this assumption yields precise quantitative results for the electron dynamics that agree with pump-probe and reflectivity experiments for pulse durations longer than 100 fs [15, 21], inconsistencies have been observed at shorter pulses for which a strong presence of out-of-equilibrium electron is expected [16, 20, 22].

To overcome the limitations originating from the overestimation of the electron energy, various revised models have been proposed based on: (i) Boltzmann's transport equations [23], (ii) three temperature models [18, 20], (iii) two temperature models with the introduction of two source terms [22, 24].

The above approaches described successfully both the ultrafast dynamics and thermal response of the irradiated material in many physical systems [17, 18, 20, 23, 25]. Nevertheless, although those methodologies appeared to illustrate efficiently the role of the nonthermal electrons in the subsequent relaxation processes (i.e. that achieve thermal equilibration between carrier and lattice through carrier-phonon coupling), some of the above models were applied only for *metals* (i.e. consideration of an infinitesimal nonthermal, steplike change of the electronic distribution due to the irradiation and promotion of electrons to the unoccupied states above the Fermi energy) [16, 17, 20, 22, 24-28]. One very intriguing challenge is whether similar models can be developed for other materials (i.e. semiconductors or dielectrics) where excitation and relaxation processes include more complex mechanisms such as multiphoton/tunnelling and impact ionisation as well as carrier recombination. It is evident that a revision to existing models is required to account for the behaviour of out-of-equilibrium carriers in the conduction band and their interaction with thermalised carriers and lattice when the processes are considered. However, validity of a simplistic extension of the aforementioned models is rather questionable. By contrast, due to the complexity of the physical mechanisms that are involved, an approach based on quantum mechanical principles is regarded as a more precise technique to describe the underlying ionisation processes and ultrafast dynamics. To address this need, simulations based on Density Functional Theory (DFT) have been applied in various systems [29, 30] and the impact of out-of-equilibrium electrons in the subsequent relaxation processes has been successfully evaluated. Nevertheless, one still not explored process in these approaches is that they do not consider potential temporal variation of the optical parameters (and therefore energy absorption) of the irradiated material induced by the presence of hot carriers, which becomes significant at extremely short pulses.

One very promising wide-band gap material is SiC and its polymorphs due to its impact to numerous technological applications. More specifically, the advantages of SiC devices are opening up for advanced applications in the most important fields of electronics while its properties allow the performance of existing semiconductor technology to be extended [31, 32]. Although the properties of this material have been widely explored, response upon extreme heating is an area that has yet to be investigated.

To address the above challenges and apply the methodology to explore physical processes after irradiation of SiC with single femtosecond laser pulses, a two-tiered approach is followed to describe two regimes: (i) a real time simulation is presented to compute the ultrafast dynamics of the *out-of-equilibrium* excited carriers as well as the induced optical parameters for hexagonal Silicon-Carbide (6H-SiC) (Section II), (ii) a revised TTM for semiconductors is employed to provide a description of the temporal evolution of the temperatures of the carriers/lattice population and recombination process for the produced *thermalised* population of excited carriers and the energy relaxation between the carriers and the lattice systems via carrier-phonon coupling. A detailed analysis of the results the theoretical model yields is presented in Section III for various values of the laser fluence while an estimation of the surface damage threshold is calculated. Concluding remarks follow in Section IV.

## II. THEORETICAL MODEL

### a. Structure of 6H-SiC

Silicon carbide is a unique material as it occurs in some 250 polymorphs. A particular kind of polymorphism which is called polytypism, occurs in certain close-packed structures: two dimensions of the basic repeating unit cell remain constant for each crystal structure while the third dimension is a variable of a common unit perpendicular to the planes with the closest packing. Polytypes consist of layers with specific stacking sequence where the atoms of each layer can be arranged in three configurations in order to maximize the density [33]. The fundamental structural unit is a covalently bonded tetrahedron of four Carbon (C) atoms with a single silicon (Si) atom at the centre. On the other hand, each C atom is surrounded by four Si atoms. Among the various polytypes of SiC, the hexagonal 6H[1] configuration is one of the most widely studied [33, 34] and it will be the focus of this work. In Fig.1, the unit cell of 6H-SiC is shown which has a complex structure with 12 atoms (Fig.1).

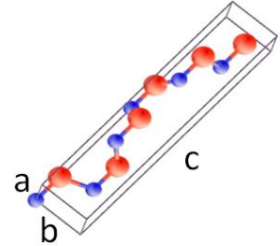

Figure 1: Structure of 6H-SiC: Silicon atoms are represented by large spheres (*in red*) corresponds while Carbon atoms are represented by small spheres (*in blue*) The cell parameters are $a=b$=3.095 Å, $c$=15.18 Å [35].

### b. First principles calculations

Polytypism has a strong influence on the material physical and chemical properties. In particular, the optical properties of SiC and their relation to the polytypic character have been extensively investigated [36-39]. These studies include measurements of the dielectric function, the refractive index, as well as the determination of the frequency-dependent dielectric function, optical absorption and reflectivity spectra and are connected with the band structure of the material. A precise evaluation, however, of the optical properties for systems in nonequilibrium states due to excitation conditions require also consideration of correlation effects, (i.e. excitonic effects due to electron-hole Coulomb interaction) or plasmons. A consistent

---

estimation of the role of excitonic effects can be derived from the solution of the Bethe–Salpeter (BS) equation for the electron–hole Green's function, within the many-body perturbation theory (MBPT) framework [36].

In this work, Yambo, was used to address the above issues [40]. Yambo is a consistent *ab initio* code for calculating quasiparticle energies [41] and optical parameters of electronic systems within the framework of MBPT. Although alternative approaches have been used, including *ab-initio* Molecular Dynamics techniques [42], the Yambo code has proven to be an efficient, well-established algorithm to describe excitation and dynamics following irradiation of materials with intense sources (see [43] and references therein). Using Yambo code, the equilibrium properties were computed starting from a self-consistent calculation of the Kohn-Sham eigenvalues and eigenstates in the DFT framework within the local-density approximation. DFT calculations were performed with the Quantum-Espresso code [44] using the Perdew-Burke-Ernzerhof functional [45] and norm–conserving pseudopotentials. Compared to other polytypes of SiC, the analysis of 6H-SiC in terms of band-to-band transitions is more demanding and complex due to the large number of bands being folded into the small Brillouin zone. A shifted 8×8×2 k-point sampling for the ground-state was used, while a kinetic energy cut-off of 100 Ry was considered. The quasiparticle corrections to the fundamental band gap have been calculated from the standard GW approximation (*G* stands for the one-body Green's function and *W* for the dynamically screened Coulomb interaction) with the Godby-Needs plasmon-pole model and applied as a rigid shift to all the bands. Calculations of the quasi-particle energies and optical susceptibilities have been performed using the Yambo code [40] and a total of 100 bands for Green's function expansion was used. The energy positions of the top of the valence band and the bottom of the conduction band are calculated along with the direct and indirect bandgaps through the Yambo code. According to the calculations, the Fermi level is estimated to be at 10.33 eV, while the energy band gaps are computed to be equal to 2.03 eV and 3.16 eV for the indirect and direct band gaps, respectively.

Following the evaluation of the ground state properties (Quantum Espresso is used to perform this step and determine the dielectric function as a function of the photon energy), the evolution of the electronic system under intense laser irradiation requires performance of real time (RT) simulations. Yambo is, subsequently, employed to compute the out-of-equilibrium carrier distribution within the pulse duration assuming the laser pulse characteristics (energy, shape, duration, polarization). This step is implemented in Yambo with a recently introduced feature that allows monitoring the real time carrier dynamics within the non-equilibrium MBPT framework. By numerically integrating the time-dependent equation of motion for the density matrix expressed in the space of the single particle wave-functions, the time evolution of the non-equilibrium carrier distribution is computed. In turn, the carrier occupations are subsequently used to monitor the time evolution of optical parameters. An essential step of the RT simulation is the removal of all symmetries since the external laser field breaks the symmetry which eventually leads to the development of polarisation effects (for a detailed description, see Ref.[43]). Hence, polarisation effects are expected to be closely related both to the photon energy and the laser fluence and it will be reflected on the excitation level of the carriers (i.e. values of the DFT-based calculated carrier densities as shown in Section III).

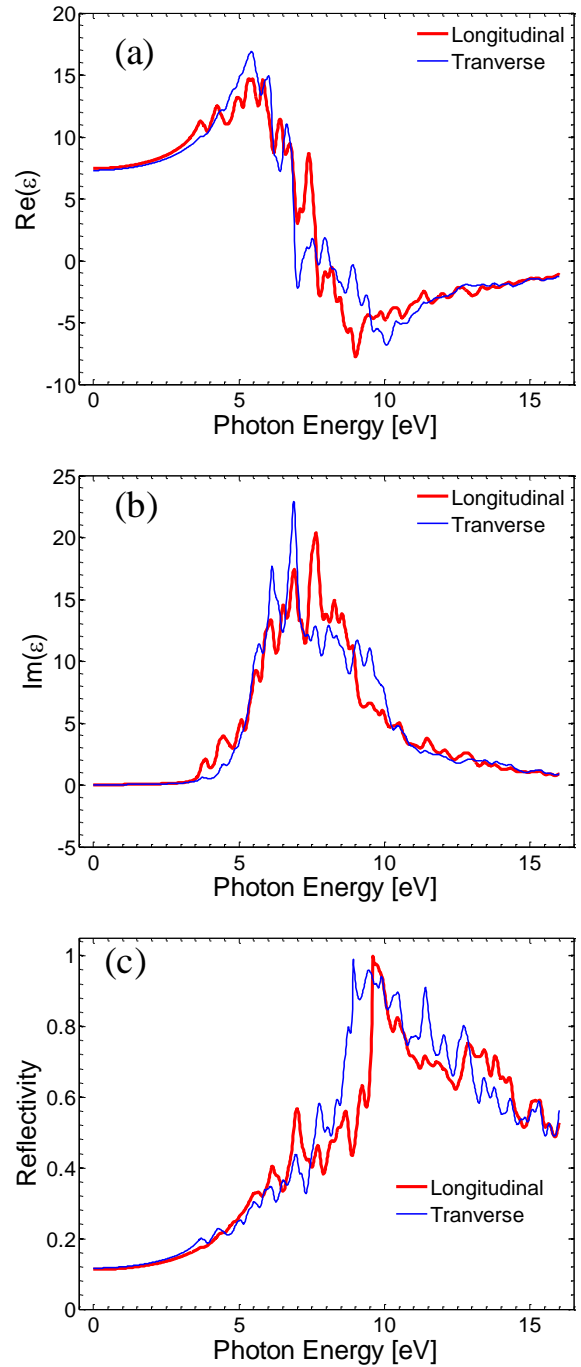

Figure 2: Simulated results for the real and imaginary part of dielectric function (a,b) and (c) reflectivity at various photon energies assuming longitudinal and transverse dielectric constant components.

In the current work, single shot laser pulses are used impinging normally to the sample. To simulate the temporal profile of the pulse, it is important to set precisely the propagation variables such as the time interval, the duration of the simulation, the integrator and the pulse intensity. In regard to the laser pulse shape and polarisation, a linearly polarized Gaussian pulse has been chosen that is centred at the fundamental absorption peak in order to generate a significant amount of carriers. The number of carriers is expected to increase as long as the pulse intensity is nonzero.

Simulation results for the optical parameters of the irradiated material are illustrated in Fig.2 for various photon energies that correspond to laser wavelengths in the range [73 nm, 12μm] at 300 K. The calculation of the optical parameters such as the refractive index $n$, extinction coefficient $k$, and reflectivity $R$ of the material can be derived from the following expressions based on the results for the dielectric constant $\varepsilon$ (Fig.2)

$$
\begin{aligned}
n &= \sqrt{\frac{|\varepsilon|^2 + \mathrm{Re}(\varepsilon)}{2}} \\
k &= \sqrt{\frac{|\varepsilon|^2 - \mathrm{Re}(\varepsilon)}{2}} \\
R &= \frac{(n-1)^2 + k^2}{(n+1)^2 + k^2}
\end{aligned}
\tag{1}
$$

while the absorption coefficient is given by $\alpha = 4\pi k/\lambda_L$ where $\lambda_L$ stands for the laser wavelength. Results show the frequency dependence of the real and imaginary parts of the dielectric function along the transverse and longitudinal directions, respectively (Fig.2a,b). A comparison with theoretical and measured values for the longitudinal component of the dielectric function $\varepsilon$ (Fig.2a,b) reported in previous works shows a remarkable agreement that illustrates the validity of the approach [36, 38, 46]. It is noted that while there is a discrepancy for $\varepsilon$ values between the two polarization directions, there is no difference in $\varepsilon$ for the transverse and longitudinal components for the photon energy used in this work (3.09 eV). An interesting aspect is the 'metallic' behaviour (i.e. $Re(\varepsilon)<0$) that is exhibited in both transverse and longitudinal spectra by the irradiated material at laser wavelengths $\lambda_L<179$ nm (photon energies larger than 6.9 eV). It is also noted that the energy absorption and its spatial attenuation during the pulse is treated via $k$ and $\alpha$ (Eq.1).

### c. Energy and Particle Balance equations

#### (i) Carrier excitation and carrier-phonon relaxation processes assuming instantaneous carrier thermalisation

To describe the carrier excitation and relaxation processes for semiconductors, the relaxation time approximation to Boltzmann's transport equation has been widely employed [19, 47-53] to determine the spatial and temporal dependence ($t$) of the carrier density number, carrier energy and lattice energy. The carrier system is assumed to be non-degenerate (i.e. Maxell-Boltzmann distributed) as the adoption of a more rigorous approach is not expected to lead to substantial differences in the evaluation of the main observable effects (i.e. damage thresholds [50]). To describe the carrier dynamics and associated thermal effects following excitation, energy and particle balance equations are used to derive the evolution of the carrier density number $N_c$, carrier temperature $T_c$ and lattice temperature $T_L$ [48, 50, 51]. More specifically, the balance equation for the lattice subsystem yields that the lattice energy density rate $\frac{\partial U_L}{\partial t}$ should be equal to $\vec{\nabla} \cdot \left( K_L \vec{\nabla} T_L \right) + g(T_c - T_L)$ in which the first term describes energy transport in the lattice system ($K_L$ is the lattice heat conductivity) and the second term describes the energy exchange of the lattice with the carrier system ($g$ stands for carrier-phonon coupling coefficient). Given that $C_L = \frac{\partial U_L}{\partial T_L}$, the following equation is derived

$$
C_L \frac{\partial T_L}{\partial t} = \vec{\nabla} \cdot \left( K_L \vec{\nabla} T_L \right) + \frac{C_c}{\tau_c}\left( T_c - T_L \right)
\tag{2}
$$

By contrast, the balance equation for the carrier system is more complicated as the carrier energy density is dependent not only on $T_c$ but also on $N_c$ and on the energy band gap $E_g$ [48, 50, 51]. Thus, the carrier energy density rate $\frac{\partial U_c}{\partial t}$ is given by the following expression

$$
\frac{\partial U_c}{\partial t} = C_c \frac{\partial T_c}{\partial t} + \frac{\partial N_c}{\partial t}\frac{\partial U_c}{\partial N_c} + \frac{\partial E_g}{\partial t}\frac{\partial U_c}{\partial E_g}
\tag{3}
$$

where $C_c$ is the carrier heat capacity ($C_c = 3N_c k_B$, $k_B$ stands for the Boltzmann constant [48, 50, 51]). On the other hand, in a non-degenerate system, $U_c$ is given by $U_c = N_c \left( E_g + 3k_B T_c \right)$ (i.e. equal to the product of the carrier number density and the sum of the band-gap energy per unit volume and the kinetic energy of the electrons and holes; the latter is equal to $2 \times 3/2 k_B T_c$ for a non-degenerate carrier system to accommodate both the electron and hole densities [47, 48, 50]) while the balance equation for the carrier system is provided by the expression $C_c \frac{\partial T_c}{\partial t} = -g(T_c - T_L) + LE_1(E_{photon}, I)$ (the first term in the second part describes the energy exchange between the carriers and the lattice system while $LE_1(E_{photon}, I)$ is dependent on the intensity of the laser $I$ and the photon energy $E_{photon}$ which is related to the energy provided to the carrier system by the laser source [47]).

The above discussion leads to the following equation for the $T_c$ rate ($g = \frac{C_c}{\tau_c}$, where $\tau_c$ is the carrier-phonon energy relaxation time [47, 48, 51, 52])

$$C_c \frac{\partial T_c}{\partial t} = -\frac{C_c}{\tau_c}(T_c - T_L) - \frac{\partial N_c}{\partial t}\left( E_g + 3k_B T_c \right) - N_c \frac{\partial E_g}{\partial t} + LE_1(E_{photon}, I) \quad (4)$$

Finally, the particle balance equation is related to the rate of the carrier density following excitation of the material and it is given by the expression

$$\frac{\partial N_c}{\partial t} = -\gamma N_c^3 + LE_2(E_{photon}, I) \quad (5)$$

where the first term in the second part is related to Auger recombination ($\gamma$ stands for the Auger recombination coefficient that leads to a gradual reduction of carrier density while $LE_2(E_{photon}, I)$ includes various excitation mechanisms such as interband and intraband absorption processes, impact ionisation, etc. [47, 48, 50, 51]). It is noted that no carrier current or heat current density is considered (in previous studies, simulations manifested that neglecting heat dissipation and particle transport are not expected to produce significant changes to the material response [47, 48, 50]). The above set of *coupled* nonlinear equations (Eqs.2,4,5) constitutes the main theoretical framework that is used to describe the carrier density and thermal evolution ($N_c$, $T_c$ and $T_L$) for a semiconducting material [47]. Despite the underlying complexity of the physical processes, the above rate equations (Eqs.2,4,5) have successfully described ultrafast phenomena and relaxation processes in a wide range of materials such as metals, semiconductors and dielectrics [13, 27, 47, 48, 50-65]. An assumption that is usually made in modelling carrier excitation and relaxation processes is that carriers are considered to *thermalize instantaneously* (i.e. a delta function equilibration is assumed) which in principle, it is true for long pulses. *Nevertheless*, as noted in the introductory section, the use of these equations is questionable if an out-of-equilibrium carrier population is formed which occurs for very short pulses (<100 fs). A detailed account of the energy of out-of-equilibrium carriers cannot be described by the balance equations presented in the previous section as a thermalization of the out-of-equilibrium carriers to a hot Fermi distribution is required. Thermalisation of the carrier system is achieved mainly through carrier-carrier scattering processes that allows formation of a Fermi distribution for the carriers with a well define electron temperature which means that before thermalisation is completed, carrier temperature is *not* defined and therefore Eqs.2-4 *cannot* be used.

### (ii) Carrier-phonon relaxation processes assuming formation of out-of-equilibrium carrier population

Therefore, in contrast to the traditional methodology of using Eqs.2,4,5 to describe energy absorption, carrier excitation carrier internal energy and densities, an alternative methodology is used through, firstly, the employment of DFT approaches presented in Section IIb.

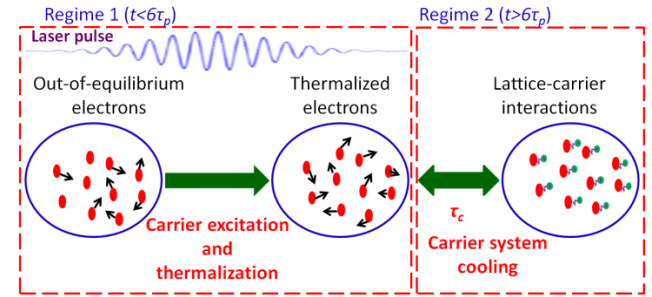

Figure 3: Processes following irradiation with ultrashort pulses: Regime 1: energy absorption, production of hot (out-if-equilibrium) carriers, carrier thermalisation, Regime 2: Carrier cooling through lattice-carrier interaction and equilibration process.

On the other hand, a rapid carrier thermalisation is assumed to have been completed at the end of the pulse. Therefore, to describe the physical processes following irradiation 6H-SiC with very short femtosecond pulses, two regimes are investigated (Fig.3): (i) Regime 1: for $t < 6\tau_p$ ($\tau_p$ stands for the laser pulse duration) where DFT calculations are performed to determine laser energy absorption and ultrafast dynamics, (ii) Regime 2: for $t > 6\tau_p$ when laser is considered to have been switched off. In the latter case, as carrier thermalisation is assumed to have been completed, a

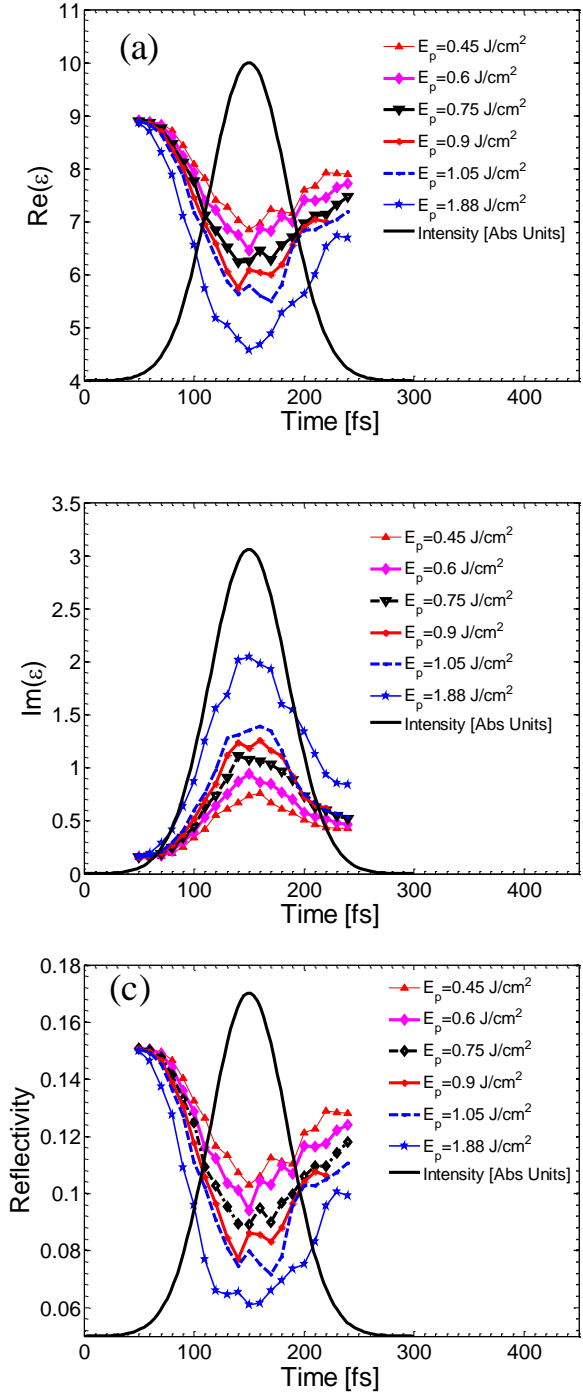

Figure 4: Evolution of (a) real part, (b) imaginary part of dielectric constant, (c) reflectivity. ($\lambda_L$ = 401 nm, $\tau_p$=50 fs). *Black solid* line shows the laser intensity profile.

*modified* version of Eqs.2,4,5 should be used to derive the evolution of the carrier densities and temperatures of the carriers and lattice. Notably, in contrast to the traditional

TTM that has a source term which gives rise to carrier excitation, an appropriate modification is required to show departure from the standard approach. More specifically, in the revised TTM model which is used in this work, it is assumed that $LE_1(E_{photon}, I)$=0 for $t$>$6\tau_p$ (as pulse has been switched off) while the carrier density value attained at the end of the pulse is derived through DFT calculations as explained in Section II; on the other hand, $N_c$ evolution after the laser pulse has been switched off takes into account feedback from those calculations (see next Section). Similarly, $LE_2(E_{photon}, I)$=0 for $t$>$6\tau_p$. Thus, the evolution of the carrier density for $t$>$6\tau_p$ is calculated through the expression

$$N_c = -\int_{t=6\tau_p}^{t} \gamma N_c^3 dt + PE(t=6\tau_p) \qquad (6)$$

after integration of Eq.5, where $PE(t$=$6\tau_p)$ is the carrier density which is computed through DFT calculations and real time simulations (see Section III). $PE$ is attributed to polarization effects contributions as emphasized in Section II. Notably, this term is not included in the traditional TTM.

On the other hand, equilibration of the thermalized carrier system with the material is performed through carrier-phonon coupling (i.e. $\frac{C_c}{\tau_c}(T_c - T_L)$ ) however, as the pulse duration is too short, it is assumed that its influence is not very significant before the laser pulse ends. Hence, the set of *coupled* set of nonlinear equations used to describe the carrier and thermal evolution of the system at $t$>$6\ \tau_p$ is the following

$$
\begin{aligned}
C_c \frac{\partial T_c}{\partial t} &= -\frac{C_c}{\tau_c}(T_c - T_L) - \frac{\partial N_c}{\partial t}\left(E_g + 3k_B T_c\right) - N_c \frac{\partial E_g}{\partial T_L}\frac{\partial T_L}{\partial t} \\
C_L \frac{\partial T_L}{\partial t} &= \vec{\nabla}\cdot\left(K_L \vec{\nabla} T_L\right) + \frac{C_c}{\tau_c}(T_c - T_L) \\
\frac{\partial N_c}{\partial t} &= -\gamma N_c^3
\end{aligned}
\qquad (7)
$$

The parameter values that are used in this work for 6H-SiC are the following: for $C_L$, the lattice heat capacity, a temperature dependent expression is derived through fitting of data in Ref.[66], $\gamma$=7 × 10$^{-31}$ cm$^6$/s [67], $\tau_c$ ~300-500 fs [48, 50, 53]) and $E_g$ that corresponds to the $T_L$ dependent energy band gap of 6H-SiC is taken to be equal to $E_g$ = 3.01 - 6.5 x 10$^{-4}$ × $(T_L)^2$ /( $T_L$ + 1200) eV [68, 69] (this is the reason why $\frac{\partial E_g}{\partial t}$ in Eq.4 turns into $\frac{\partial E_g}{\partial T_L}\frac{\partial T_L}{\partial t}$ in the first equation in Eq.7). The aforementioned expression is used to provide the evolution of the band gap for $t$>$6\tau_p$ where a lattice temperature gradient occurs that could lead to $E_g$ shrinkage. In principle, the temperature effect on the energy bands of the semiconductor and hence the band gap of the material is a cumulative effect of thermal lattice expansion and electron-phonon interaction. On the other hand, as noted in the previous section, at smaller timepoints $t < 6\tau_p$ and during the laser-based excitation time, $E_g$ is calculated from the Yambo code and it corresponds to the difference between the top and bottom energy positions of the valence and conduction bands, respectively. It is noted that the direct energy gap (equal to 3.16 eV as computed in Section II) is used for $E_g$ in $t < 6\tau_p$ as the indirect band gap corresponds to a phonon-assisted interband transition that is less likely to occur given the insignificant phonon system energy during the pulse. It is also emphasised, though, that in this work, *single* shot simulated experiments were performed in which disorder or lattice deformation due to laser irradiation were not considered. It is evident that in multiple shot conditions a more precise description of the response of the material should be obtained by taking into account the role of defects or disorder induced by laser heating in the calculations of the energy gap. On the other hand, for $t > 6\tau_p$, although remarkable changes to the thermal response of the system are not expected, a rigorous approach through the aforementioned temperature-dependent expression ($E_g = 3.01 - 6.5 \times 10^{-4} \times (T_L)^2 / (T_L + 1200)$ eV) is used in this work. It is noted that thermally generated effects (i.e. thermal expansion, strain propagation, plastic deformation, defect formation, etc.) can also be developed as a result of the irradiation with intense femtosecond pulses ([24, 57, 65, 70]), however, such an investigation is beyond the scope of this work.

Finally, $K_L = 611/(T_L - 115)$ Wcm$^{-1}$K$^{-1}$ [66]. In previous works, an anisotropic heat conductivity was reported for various polytypes of SiC including 6H-SiC in which it was shown that the cross-plane thermal conductivity $K_L^{(z)}$ (perpendicular to the hexagonal planes) of 6H-SiC is 30% lower than its in-plane thermal conductivity $K_L^{(z)}$ (parallel to the hexagonal planes) [71, 72]. Experimental observations indicate that the anisotropy in the thermal conductivity 6H- SiC is expected due to the hexagonal Bravais lattice structure which suggests that in general this difference should not be ignored in a rigorous investigation. Nevertheless, in this study, it is assumed that for the computation of the damage thresholds, results are not expected to be remarkably sensitive to the 3D character of the heat diffusion. Certainly, a more precise exploration of the impact of the anisotropy on thermal effects could provide a more detailed account of the role of directional heat diffusivity, however, this investigation is beyond the scope of the present study. Therefore, for the sake of simplicity, a bulk material is considered while the laser spot radius is taken to be substantially larger than the thickness of the material and, thereby, a 1-d solution is considered to sufficiently determine the carrier dynamics and thermal response of the system.

## III.  RESULTS  AND  DISCUSSION

A quantitative description of carrier excitation, relaxation processes, and thermal response of both the carrier and lattice systems is provided through the use of the aforementioned DFT+TTM combined model. To highlight the contribution of the nonthermal carriers to the transient dynamics of the system, femtosecond pulsed laser beams of duration, significantly smaller than the carrier-phonon energy relaxation time, are assumed ($\tau_p$=50 fs). The photon energy of the laser beam is $\hbar\omega = 3.09$ eV which corresponds to laser beam wavelength $\lambda_L$=401 nm and it is similar to the size of the material's computed energy band gap ($\cong$3.16 eV). The initial conditions are $T_e(t=0) = T_L(t=0) = 300$ K, and $N_e =10^{12}$ cm$^{-3}$ at $t$=0. The (peak) fluence is equal to $E_p = \sqrt{\pi}\tau_p I_0 / \left(2\sqrt{ln2}\right)$, where $I_0$ stands for the peak intensity.

The optical parameters evolution (real and imaginary part of the dielectric function, and reflectivity $R$) are illustrated in Fig.4 for six various fluence (peak) values, 0.45, 0.6, 0.75, 0.9, 1.05, and 1.88 Jcm$^{-2}$. For all fluence values, DFT calculations showed a decreasing reflectivity reaching a minimum at $t$=6$\tau_p$ before a relaxation to the initial reflectivity value ($\varepsilon$ ($\lambda_L$= 401 nm) $\cong$ 8.9+0.15$i$ shown also in Fig.2). This behavior resembles that demonstrated by lower band-gap semiconductors upon irradiation with laser pulses of duration that is comparable with $\tau_c$, and fluences that are not high enough to induce 'metallisation' ($Re$ ($\varepsilon$)<0 [52, 53]) of the irradiated material. Interestingly, both for the fluences used in this work as well as for even larger values which correspond to intensities where the material appears to undergo a phase transformation or even ablation, $Re(\varepsilon)$ never becomes negative (Fig.4a). On the other hand, a noticeable variation of the imaginary part of the dielectric function $Im(\varepsilon)$ is predicted (Fig.4b) that is also related to the free electron absorption coefficient and significant response of the excited electron system. Furthermore, transient reflectivity calculations (Fig.4c) illustrate a substantially large drop during the pulse duration that further increases the laser energy absorption. By contrast, larger laser energies allow increase of excitation at larger depths (Fig.4b).

To quantify the carrier population in the simulations, it is noted that the volume of the unit cell equals 838.1109 atomic units that corresponds to 1.23368 × 10$^{-22}$ cm$^{-3}$. The carrier density evolution illustrated in Fig.4a that result from DFT calculations indicates an initial increase of the carrier population that reaches a peak value where laser intensity is higher before a sharp decrease occurs. The temporal decrease of reflectivity (Fig.4) and, thereby, increase of the absorbed laser energy is projected on the increase of excited carrier density as higher excitation conditions are induced (Fig.5a). It is noted that, the initial decrease (from a peak value) of carrier density that is shown in Fig.5a is due to some kind of polarisation effects. These effects are usually small at resonances but they become more important outside the resonance regimes.

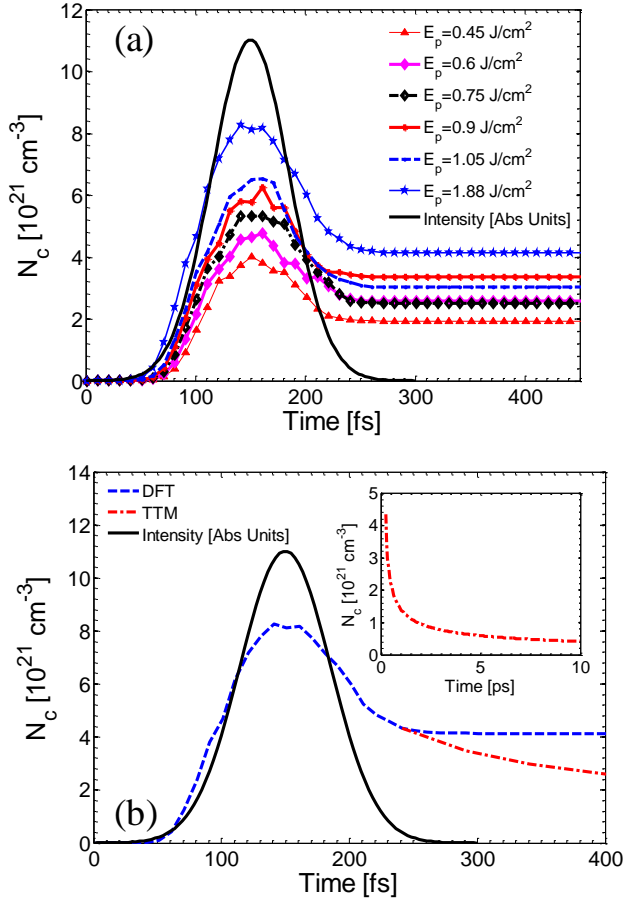

Figure 5: Evolution of carrier density (a) through DFT calculations, (b) results for $E_p$=1.88 J/cm² ($\lambda_L$= 401 nm, $\tau_p$=50 fs). Inset illustrates the carrier evolution at larger timepoints through Eq.5. *Black solid* line shows the laser intensity profile.

Similar behaviour has been reported in previous works in which the decrease of carrier density is attributed to recombination effects [30]. Variation of the carrier density values (Fig.5) with fluence demonstrates also that the laser energy affects the induced polarisation effects. As noted in Section II, similarly, there is also a dependence of the polarisation effects on photon energy as excitation levels at different laser wavelengths is also expected to change.

Notably, DFT calculations demonstrate that after the end of the pulse, the carrier density evolution remains constant unlike an anticipated decrease predicted in other semiconductors in different irradiation conditions [47, 50, 52, 53]. The absence of a further decreasing behaviour is due to the fact that (Auger or radiative) recombination processes are not included in the DFT model. Certainly, the incorporation of such processes in the quantum mechanical approach would allow a more precise description of carrier transient evolution. Recombination and other scattering processes could be introduced by selecting appropriate approximations for the self-energy and introducing a dynamical character to the self-energy. These additions, though, would make the approach more demanding which is beyond the scope of the present study [73].

On the other hand, the model presented in this work is aimed to combine the DFT-based calculations and TTM results by linking the description in the two different regimes where *out-of-equilibrium* (Regime 1) and *thermalised* (Regime 2) carriers are present. Therefore, to allow an efficient description of carrier dynamics, some physically consistent methodology is required to link the two regimes. To correlate the carrier temperature of a thermalized population with their density, it is assumed that at the end of the pulse, the carriers have reached their maximum thermal energy and maximum carrier temperature $T_c^{\max}$ and that after that moment, they stop to receive energy from the laser source (while $T_c$ starts to drop due to carrier-phonon scattering processes) [47, 48, 52]. Furthermore, it is assumed that at the end of the pulse, carriers have thermalised and a Fermi-Dirac distribution with a well-defined temperature has been re-established [15].

Given the anticipated insignificant variation of the lattice temperature within the pulse duration due to the small $\tau_p$ and the large heat capacity of the lattice system for semiconductors compared to $C_c$, $T_L$ is approximately equal to $T_L^0 = 300$ K at the end of pulse. It is noted that in other materials such as metals with smaller heat capacity, hot electron-phonon scattering processes lead to a rather significant increase of the lattice temperature within the pulse duration [24, 26]. By contrast, similar notable increase of $T_L$ is not expected for 6H-SiC. However, a more thorough investigation that provides a more conclusive estimation of the lattice temperature is beyond the scope of the present work.

To solve Eqs.7, an explicit forward time centered space finite difference scheme is used [53]. While Eqs.7 can be also employed to provide a 3-D solution assuming the spatial characteristics of the beam profile, for the sake of simplicity, and for the objectives of the present study, equations are solved in 1-d, along a line between $z$=0 μm and $z$=5 μm. It is assumed that on the boundaries, von Neumann boundary conditions are satisfied and heat losses at the front and back surfaces of the material are negligible. A common approach followed to solve similar problems is through the employment of a staggered grid finite difference method which is found to be effective in suppressing numerical oscillations. Temperatures ($T_c$ and $T_L$) and carrier densities ($N_c$) are computed at the centre of each element while time derivatives of the displacements and first-order spatial derivative terms are evaluated at locations midway between consecutive grid points [51].

Starting from the values of $N_c$ calculated from the DFT approach (Fig.5a), a correction to the carrier density evolution is required to account for Auger recombination. More specifically, the third equation of Eqs.7 is used to produce the rate of the carrier density for $t>6\tau_p$ while the initial carrier density to derive $T_c^{max}$ corresponds to the value of $N_c$ at $t=6\tau_p$ for which DFT calculations predict a carrier density that stops to decrease further. This value is taken to be the contribution of polarisation effect at $t=6\tau_p$ (i.e. $PE(t=\tau_p)$ in Eq.6).

Considering the above assumptions, the maximum carrier temperature $T_c^{max}$ is calculated by the first equation of Eqs.7 assuming $\frac{\partial T_c}{\partial t}=0$. Then, Eqs.6,7 lead to the following expression for $T_c^{max}$

$$T_c^{max} \cong \frac{T_L^0\left(\frac{C_c}{\tau_c}+\frac{N_cC_c}{C_L\tau_c}\frac{\partial E_g}{\partial T_L}\right)-\gamma N_c^3 E_g}{\frac{C_c}{\tau_c}+\frac{N_cC_c}{C_L\tau_c}\frac{\partial E_g}{\partial T_L}-3k_B\gamma N_c^3} \quad (8)$$

Eqs.7-8 allow the calculation of the evolution of the carrier densities (including the correction due to Auger recombination), as well as the temporal dependence of the carrier and lattice. Results and correction to the carrier density evolution profile are shown in Fig.5b for 1.88 Jcm$^{-2}$ (similar behavior is predicted for other fluences) while the inset depicts the transient dynamics of $N_c$ at larger timepoints. In Fig.5b, the *blue* dashed line is derived from DFT calculations (see also Fig.4a) while the *red* dashed-dotted line results from the use of Eqs.7-8. Notably, the significant decrease of $N_c$ resulting from the contribution of recombination processes is manifestly illustrated in Fig.5b which indicates the Auger recombination role should not be ignored. The significance of Auger recombination in both the carrier dynamics [74] and surface modification processes have been also revealed in previous reports [75].

On the other hand, the thermal response of the carrier and lattice system on the surface of the material for laser fluence equal to 1.88 Jcm$^{-2}$ is summarised in Fig.6. It is evident that a maximum carrier temperature occurs at $t=6\tau_p$ that is subsequently followed by a decrease due to carrier-lattice heat transfer and relaxation of the system. Relaxation processes and exchange of energy between the carrier and lattice subsystems yield a similar behaviour to what occurs in other materials [47, 50, 52, 53]. Furthermore, the simulated maximum $T_L$ values allow an estimation of the damage threshold of the material (~1.88 Jcm$^{-2}$). It is noted that, in this work, damage threshold is associated to the fluence value at which the surface lattice temperature exceeds the melting point of the material [24, 51, 57, 76] ($T_{melting}$=3100 K for 6H-SiC [77]).

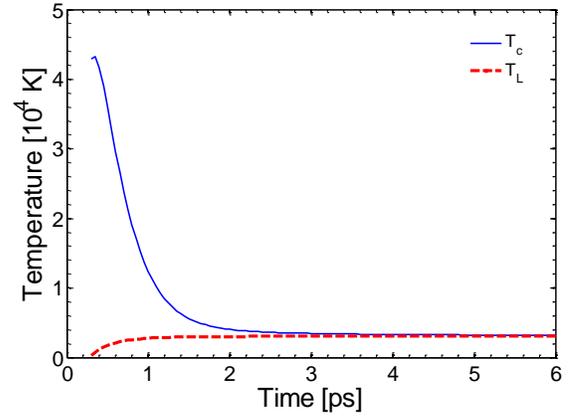

Figure 6: Evolution of electron and lattice temperature ($\lambda_L$= 401 nm, $\tau_p$=50 fs, $E_p$=1.88 J/cm²).

Certainly, the aforementioned methodology and predictions that are used to provide an estimate for the damage thresholds require validation of the model with experimental results. To the best of our knowledge, there are not similar reports with experimental observations for the pulse duration and laser wavelengths considered in this work. Nevertheless, experimental measurements for damage thresholds illustrated in Fig.7 at various laser wavelengths and pulse durations indicate that the theoretical value for the critical fluence used for the simulations conditions in this work represents a reasonable prediction: experimental measurements at various wavelengths (Fig.7) show a dispersion of the damage threshold estimations while the simulated value appears to be within the range of the measured values [78-82]. Certainly, other effects should also be taken into account to provide a conclusive picture such as reflectivity changes at different wavelengths (Fig.2c) and role of multiphoton absorption.

On the other hand, there is a number of reports about laser induced periodic surface structures (LIPSS) which are formed on 6H-SiC crystal irradiated by femtosecond laser pulses at various wavelengths [83]. Experimental results for irradiation with multi-shot laser pulses at 400 nm indicate a measured fluence threshold for LIPSS formation which is approximately equal to 0.49 J/cm² [83] while the model yields a fluence threshold approximately equal to 1.88 J/cm² for single shot simulations. Similarly, bulk ablation of 6H-SiC at 785 nm takes place at a fluence of 1.4 J/cm² [84] while nanoripples have lower damage threshold than bulk single crystals which has also been observed in other studies [83]. A possible reason can be attributed to the fact that, in principle, an experimentally observed formation of LIPSS requires irradiation with multiple number of pulses ($NP$>10 shots [85]). By contrast, it is known that in transparent materials and semiconductors [86, 87], the damage threshold for surface modification at increasing $NP$ drops significantly (more than ¼ of the value for $NP$=1) compared to the measured value for single shot experiments due to the presence of defects and incubation. This is expected to provide a satisfactory agreement between the predicted single laser shot-based result with the measured value (i.e. deduction of a predicted multi-shot damage threshold around 0.4 J/cm² which appears to agree with the experimental value.

A challenging issue is whether the approach followed in this work can be used to describe LIPSS formation mechanisms upon 6H-SiC irradiation with very short femtosecond pulses ($\tau_p$<100 fs). One physical process that is directly linked with the formation of periodic structures on solids is the interference of laser pulses with Surface Plasmon waves (SP) that are excited as a result of laser irradiation [49, 51, 88]. On the other hand, according to well-established theories excitation of SP requires carrier densities that lead to $Re(\varepsilon)<-1$. However, according to the simulation results in Fig.4a, despite the large decrease of $Re(\varepsilon)$ for 1.88 Jcm⁻², this parameter does not drop to sufficiently low values that can induce SP excitation despite the extremely high carrier densities which are produced (~8×10²¹ cm⁻³). This can be attributed, firstly, to the need to revise the dispersion relation that is required for SP excitation [51, 53, 89]; more specifically, the Drude-model-based dielectric function expression differs if nonthermal contributions are included that indicates that appropriate corrections have to be included. Therefore, the carrier densities evaluation for which $Re(\varepsilon)<-1$ is expected not to incubation effects, the precise role of defects in multipulse experiments (that lead to SP excitation and LIPSS formation [51, 62, 90]) and the variation caused to an effective dielectric constant should be also taken into account. These are some issues that need to be elaborated on to determine the contribution of hot electrons in incubation-related processes and surface modification mechanisms.

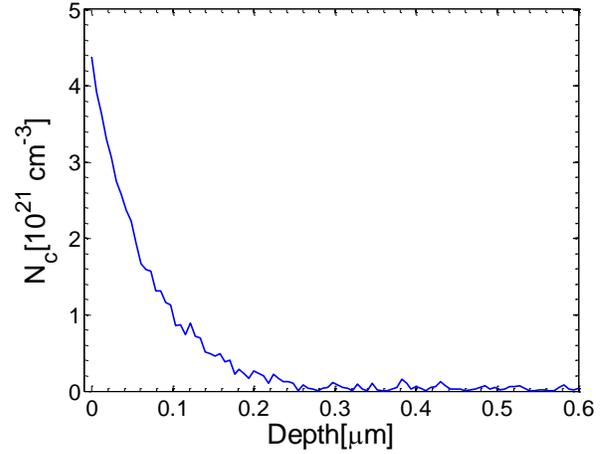

Figure 8: Spatial distribution of carrier density at $t$=300 fs ($\lambda_L$= 401 nm, $\tau_p$=50 fs, $E_p$=1.88 J/cm²).

Certainly, a more accurate conclusion will be drawn if more appropriately developed experimental (for example, time-resolved experimental) protocols are also introduced to evaluate the damage thresholds at the onset of the phase transition; similarly, pump-probe experiments could be used to validate the reflectivity changes. Furthermore, the aforementioned potential impact of anisotropy-related effects on damage thresholds should be further explored. Anisotropies in visible pump probe experiments have been previously reported by pumping at 800 nm [91].

Although results in Fig.6 are shown for irradiation with laser pulses of a single wavelength (at 401 nm), the methodology can be generalised at other frequencies for which a laser wavelength-dependent carrier is expected. More specifically, other values of the photon energies are expected to influence the energy absorption, optical parameters (Fig.2) and maximum excited carrier densities that, in turn, lead to a variation of $T_c^{max}$. Similar results have been predicted at larger wavelengths, longer pulse durations and various fluence values in a wide range of materials [13, 27, 47, 48, 50-65]. Nevertheless, extension of the investigation of the thermal response of the material following irradiation with very short pulses of different photon energy is beyond the scope of the present study.

It is, also, noted that results illustrated in this work aimed to underline the response of the system in laser conditions that lead to surface damage and therefore, be the correct condition to determine the onset of SP excitation. Secondly, given the significance of the special emphasis was given to $N_c$, $T_c$ and $T_L$ values at $z$=0. As expected, the attenuation of the laser beam energy inside the material is expected to lead to a gradual decrease of the carrier density. Results in Fig.8 illustrate the spatial distribution of the maximum carrier density at $t$=6$\tau_p$ for $E_p$=1.88 J/cm$^2$ and $\lambda_L$=401 nm up to 600 nm below the surface of the material. It is evident that at this wavelength the absorption coefficient reaches values up to 15×10$^4$ cm$^{-1}$ (Fig.4a,b). Therefore, irradiation of 6H-SiC with intense femtosecond pulses leads to large values for the absorption coefficient that is characteristic to materials that show a metallic behavior [92].

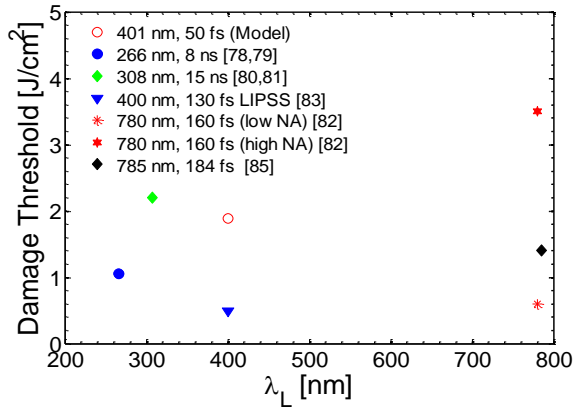

Figure 7: Threshold fluences in various conditions: experimental measurements and prediction of RT+TTM model ($NA$ corresponds to the numerical aperture of the lens used in the experiment [82]).

One aspect that is of paramount importance, is whether the above methodology can also be used to cover a wider range of potential photon energies extending to 100 eV (i.e. wavelength ~12 nm). The latter corresponds to a spectral region in which Free Electron Lasers (FEL) can be used to enable unique ultrafast scientific research [93]. At the same time, the unique output characteristics of X-ray FEL present severe requirements on the optics used to guide and shape the X-ray pulses, and the detectors used to characterize them [93, 94]. The limitation, though, that is raised in regard to the employment of the presented DFT-based methodology is that the pseudopotential which is used for the calculations in this work assumes that core-electrons are not excited. This assumption at substantially larger photon energies rather leads to an underestimation of the excitation levels which might be also reflected in the response of the material. Therefore, the investigation of the optical parameter values of the irradiated material at higher energies requires a revised and more precise expression of the pseudopotential that is beyond the scope of the current work.

Certainly, several parameters including a more rigorous description of the thermalisation process of the carriers, influence of scattering processes, microscopic analysis of non-equilibrium phase transition mechanisms through the use of hybrid Molecular Dynamics-TTM models [29, 95, 96] and a complete parametric investigation of the ultrafast dynamics and relaxation processes at a large range of photon energies and pulse durations should be considered towards providing a complete picture of the ultrafast processes. Nevertheless, the aforementioned framework is designed to provide, for the first time, a satisfactory methodology to link processes at two very small timescales (some hundreds of fs).

## IV. CONCLUSIONS

A theoretical framework was presented that describes both the ultrafast dynamics and thermal response following irradiation of 6H-SiC with ultrashort pulsed lasers of duration that is too short to assume an instantaneous thermalisation of excited carriers. The dynamics of produced out-of-equilibrium carrier population and thermalisation process is described through a quantum mechanical approach and real time simulations. Equilibration of the thermalised carrier system with the lattice through carrier-phonon scattering processes are presented via a revision of the classical TTM that allows the prediction of the decrease of the carrier density which is not appropriately accounted for in real time simulations. Results predict the temporal variation of the optical parameters and allow an estimation of the surface damage threshold. The theoretical framework is expected to enable a systematic analysis of the impact of the yet unexplored hot carriers on surface (or even structural effects) on semiconductors through a combined DFT+TTM methodology. Predictions resulting from the above theoretical approach demonstrate that elucidating ultrafast phenomena in the interaction of matter with very short pulses (<100 fs) can potentially set the basis for the development of new tools for non-linear optics and photonics for a large range of applications.

## ACKNOWLEDGEMENTS

The authors acknowledge financial support from *Nanoscience Foundries and Fine Analysis (NFFA)–Europe* H2020-INFRAIA-2014-2015 (under Grant agreement No 654360), *HELLAS-CH project* (MIS 5002735), implemented under the "Action for Strengthening Research and Innovation Infrastructures," funded by the Operational Programme "Competitiveness, Entrepreneurship and Innovation" and co-financed by Greece and the EU (European Regional Development Fund), and COST Action *TUMIEE* (supported by COST-European Cooperation in Science and Technology). We would also like to acknowledge fruitful discussions with Davide Sangalli and Andrea Marini.

\* Corresponding authors: tsibidis@iesl.forth.gr